\documentstyle[aps,12pt]{revtex}
\begin {document}
\draft

\title{Anisotropic harmonic oscillator in a static electromagnetic
 field\thanks{published in Commun. Theor. Phys. \textbf {38} (2002)
 667-674.}}
\author{Qiong-Gui Lin\thanks{E-mail:
        qg\_lin@163.net}}
\address{Department of Physics, Sun Yat-Sen University, Guangzhou
        510275,\\
        People's  Republic of China\\
        and\\
        China Center of Advanced Science and Technology (World
        Laboratory),\\
        P.O.Box 8730, Beijing 100080, People's Republic of China
        \thanks{not for correspondence}}
\maketitle
\vskip 0.4cm
\begin{abstract}
\baselineskip 15pt {\normalsize A nonrelativistic charged particle
moving in an anisotropic harmonic oscillator potential plus a
homogeneous static electromagnetic field is studied. Several
configurations of the electromagnetic field are considered. The
Schr\"odinger equation is solved analytically in most of the cases.
The energy levels and wave functions are obtained explicitly. In some
of the cases, the ground state obtained is not a minimum wave packet,
though it is of the Gaussian type. Coherent and squeezed states and
their time evolution are discussed in detail.}
\end{abstract}

\begin{flushleft}
\baselineskip 15pt PACS numbers: 03.65.Ge, 42.50.Dv \\Keywords:
anisotropic harmonic oscillator; electromagnetic field; exact
solutions; coherent states; squeezed states
\end{flushleft}

\baselineskip 15pt
\section{Introduction}               

The anisotropic harmonic oscillator potential is of physical interest
in quantum mechanics, since it may describe the motion of, say, an
electron in an anisotropic metal lattice. With external
electromagnetic fields imposed on, such models may also be useful in
semiconductor physics \cite{01,02}. For an isotropic harmonic
oscillator potential and a homogeneous static magnetic field the
Schr\"odinger equation can be easily solved in the cylindrical
coordinates. If the harmonic oscillator potential is anisotropic, the
problem is not so easy. Nevertheless, it can be solved analytically
because the Hamiltonian is quadratic in the canonical coordinates and
momenta. There exists quite general formalisms for solving systems
with such quadratic Hamiltonians \cite{1}, and the above problem
represents a typical example \cite{2}. In Ref. \cite{2} only the
isotropic case is considered as an example for illustrating the
general method. The anisotropic case with a homogeneous static
magnetic field was studied in Refs. \cite{3,4}. In these works the
magnetic field was arranged to point in one of the rectangular axis
directions. In this paper we will solve the problem in a somewhat
easier formalism. A homogeneous static electric field is also
included, though its effect is trivial. We will also deal with a case
where the magnetic field has a more general direction. This seems not
having been considered before. Coherent and squeezed states of these
systems will be discussed in some detail. In terms of these concepts
the time evolution of some wave packets can be discussed very
conveniently.

In the next section we consider a charged harmonic oscillator with
arbitrary frequencies $\omega_x$, $\omega_y$ and $\omega_z$ in three
rectangular axis directions. A homogeneous static magnetic field in
the $z$ direction and a homogeneous static electric field in an
arbitrary direction are imposed to the system. The problem can be
easily reduced to one equivalent to the case without an electric
field by making some coordinate transformation. In Sec. III we
diagonalize the reduced Hamiltonian and give the energy eigenvalues.
Although this has been previously done, the method used here is
different and seems more convenient, especially for the subsequent
discussions of coherent and squeezed states. The wave functions are
explicitly worked out in Sec. IV. In Sec. V we discuss some special
cases, some of which may need different handling, or even cannot be
solved. In Sec. VI we consider a somewhat different case where
$\omega_z=0$ and $E_z=0$ ($E_z$ is the $z$ component of the electric
field), but the magnetic field has a $x$ or $y$ component, in
addition to the $z$ component. To our knowledge, this was not
considered before. It is solved analytically by the formalism of Sec.
III. In Sec. VII we discuss the coherent and squeezed states and
their time evolution in detail. A brief summary is given in Sec.
VIII.

\section{Reduction of the Hamiltonian}   

Consider a charged particle with charge $q$ and mass $M$, moving in
an anisotropic harmonic oscillator potential, a homogeneous static
magnetic field ${\mathbf B}=B{\mathbf e}_z$, and a homogeneous static
electric field ${\mathbf E}=E_x {\mathbf e}_x+E_y {\mathbf e}_y +E_z
{\mathbf e}_z$, where ${\mathbf e}_x$, ${\mathbf e}_y$ and ${\mathbf
e}_z$ are unit vectors of the rectangular coordinate system, and $B$,
$E_x$, $E_y$ and $E_z$ are all constants. The stationary
Schr\"odinger equation is
\begin{equation}\label{1}
H^{\rm T}\Psi({\mathbf x})={\cal E}^{\rm T}\Psi({\mathbf x}),
\end{equation}
where the Hamiltonian is
\begin{equation}\label{2}
H^{\rm T}={1 \over 2M}\left({\mathbf p}-{q\over c}{\mathbf
A}\right)^2 +{1\over 2}M(\omega_x^2 x^2 +\omega_y^2 y^2 +\omega_z^2
z^2) -q{\mathbf E}\cdot{\mathbf x},
\end{equation}
where ${\mathbf E}\cdot{\mathbf x}=E_x x+E_y y +E_z z$. Since the
magnetic field points in the $z$ direction, one may take $A_x$ and
$A_y$ independent of $z$, and $A_z=0$. Then the Hamiltonian can be
decomposed as $H^{\rm T}=H^{xy}+H^z$, where
\begin{equation}\label{3}
H^{xy}={1 \over 2M}\left[\left(p_x-{q\over c}A_x\right)^2
+\left(p_y-{q\over c}A_y\right)^2\right] +{1\over 2}M(\omega_x^2 x^2
+\omega_y^2 y^2)-q(E_x x+E_y y),
\end{equation}
and
\begin{equation}\label{4}
H^z={1 \over 2M}p_z^2+{1\over 2}M\omega_z^2 z^2-qE_z z.
\end{equation}
Therefore the total wave function can be factorized as $\Psi({\mathbf
x}) =\psi(x,y) Z(z)$, where $\psi(x,y)$ and $Z(z)$ satisfy the
following equations:
\begin{equation}\label{5}
H^{xy}\psi(x,y)={\cal E}^{xy}\psi(x,y),
\end{equation}
\begin{equation}\label{6}
H^z Z(z)={\cal E}^z Z(z),
\end{equation}
with ${\cal E}^{xy}+{\cal E}^z={\cal E}^{\rm T}$. Obviously, $H_z$
can be recast in the form
\begin{equation}\label{7}
H^z={1 \over 2M}p_z^2+{1\over 2}M\omega_z^2 (z-z_0)^2-{q^2 E_z^2
\over 2M\omega_z^2}
\end{equation}
where $z_0=qE_z/M\omega_z^2$, so that Eq. (\ref{6}) can be easily
solved with the following results.
\begin{equation}\label{8}
{\cal E}_{n_3}^z=\left(n_3+\frac 12\right)\hbar\omega_z-{q^2 E_z^2
\over 2M\omega_z^2}, \quad n_3=0,1,2,\ldots,
\end{equation}
\begin{equation}\label{9}
Z_{n_3}(z)=\psi_{n_3}^{\omega_z}(z-z_0),\quad n_3=0,1,2,\ldots,
\end{equation}
where we have quoted the standard wave functions for harmonic
oscillators:
\begin{equation}\label{10}
\psi_n^\omega(x)=\left({\sqrt{M\omega}\over 2^n n!\sqrt{\pi\hbar}}
\right)^{1/2}\exp\left(-\frac {M\omega}{2\hbar} x^2\right)
H_n\left(\sqrt{M\omega\over\hbar} x\right),\quad n=0,1,2,\ldots.
\end{equation}
Therefore our main task is to solve Eq. (\ref{5}). We take the gauge
\begin{equation}\label{11}
A_x=-{\textstyle\frac 12}B\tilde y,\quad A_y={\textstyle\frac
12}B\tilde x,
\end{equation}
where $\tilde x=x-x_0$, $\tilde y=y-y_0$, and $x_0=qE_x/M\omega_x^2$,
$y_0=qE_y/M\omega_y^2$, then $H^{xy}$ can be written as
\begin{equation}\label{12}
H^{xy}={1 \over 2M}(p_x^2 +p_y^2) +{1\over 2}M\omega_1^2 \tilde x^2
+{1\over 2}M\omega_2^2 \tilde y^2-\omega_B\tilde L_z-{q^2 E_x^2 \over
2M\omega_x^2} -{q^2 E_y^2 \over 2M\omega_y^2},
\end{equation}
where $\tilde L_z=\tilde xp_y-\tilde yp_x$, $\omega_B=qB/2Mc$, and
\begin{equation}\label{13}
\omega_1=\sqrt{\omega_x^2+\omega_B^2}, \quad
\omega_2=\sqrt{\omega_y^2+\omega_B^2}.
\end{equation}
Note that $\omega_B$ may be either positive or negative, depending on
the signs of $q$ and $B$. Now Eq. (\ref{5}) can be written as
\begin{mathletters}\label{14}
\begin{equation}\label{14a}
H\psi(x,y)={\cal E}\psi(x,y),
\end{equation}
where
\begin{equation}\label{14b}
H={1 \over 2M}(p_x^2 +p_y^2) +{1\over 2}M\omega_1^2 \tilde x^2
+{1\over 2}M\omega_2^2 \tilde y^2-\omega_B\tilde L_z,
\end{equation}
\end{mathletters}
\begin{equation}\label{15}
{\cal E}={\cal E}^{xy}+{q^2 E_x^2 \over 2M\omega_x^2}+{q^2 E_y^2
\over 2M\omega_y^2}.
\end{equation}
Eq. (\ref{14}) has the same form as that for an anisotropic harmonic
oscillator moving on the $xy$ plane under the influence of a
homogeneous static magnetic field perpendicular to the plane, except
that $x$, $y$ are replaced by $\tilde x$, $\tilde y$. The main
feature is that the reduced Hamiltonian is quadratic in the canonical
variables. Though this equation has been studied by some authors, we
will solve it in a somewhat different and simpler way in the next two
sections. Here we point out that the calculations below will be
simpler if one take the gauge $A_x=-B\tilde y$, $A_y=0$ or $A_x=0$,
$A_y=B\tilde x$. However, we prefer the above gauge (\ref{11}), since
in this gauge it would be convenient to compare the results with
known ones when $\omega_x=\omega_y$, or $\omega_1=\omega_2$.

\section{Diagonalization of the reduced Hamiltonian}

If $\omega_x=\omega_y$, Eq. (\ref{14}) can be easily solved in the
cylindrical coordinates. In the general case this does not work, so
other methods are necessary. We define a column vector
\begin{equation}\label{16}
X=(\tilde x, p_x, \tilde y, p_y)^\tau,
\end{equation}
where $\tau$ denotes matrix transposition, then the reduced
Hamiltonian (\ref{14b}) can be written as
\begin{equation}\label{17}
H={\textstyle\frac 12}X^\tau{\cal H}X ={\textstyle\frac 12}
X^\dagger{\cal H}X,
\end{equation}
where ${\cal H}$ is a c-number matrix which we take to be symmetric.
We do not write down ${\cal H}$ here since this is easy. If one can
diagonalize ${\cal H}$ then the Hamiltonian would become a sum of two
one-dimensional Hamiltonians for harmonic oscillators or something
similar (repulsive harmonic oscillators or free particles). However,
a crucial point here is to preserve the commutation relation
\begin{equation}\label{18}
[X_\alpha, X_\beta]=-\hbar(\Sigma_y)_{\alpha\beta}, \quad \alpha,
\beta=1,2,3,4
\end{equation}
after the linear transformation that diagonalizes ${\cal H}$, where
$\Sigma_y={\rm diag}(\sigma_y,\sigma_y)$, and $\sigma_y$ is the
second Pauli matrix. Therefore the needed transformations are the so
called symplectic ones \cite{1,2}. Here we will transform the
Hamiltonian into a form expressed in terms of raising and lowering
operators, so the formalism is somewhat different. Let us proceed as
follows.

Because $H$ is quadratic in the canonical variables $X$, it is easy
to show that
\begin{equation}\label{19}
[{\mathrm i}H,X]=\hbar\Omega X,
\end{equation}
where
\begin{equation}\label{20}
\Omega={\mathrm i}\Sigma_y{\cal H}
\end{equation}
is another c-number matrix. As both ${\cal H}$ and $i\Sigma_y$ are
real, so is $\Omega$. Our first step is to diagonalize $\Omega$, so
we write down it here.
\begin{equation}\label{21}
\Omega=\left(
\begin{array}{cccc}
  0 & 1/M & \omega_B & 0 \\
  -M\omega_1^2 & 0 & 0 & \omega_B \\
  -\omega_B & 0 & 0 & 1/M \\
  0 & -\omega_B & -M\omega_2^2 & 0
\end{array}
\right).
\end{equation}
The characteristic polynomial for this matrix is
\begin{mathletters}\label{22}
\begin{equation}\label{22a}
\det(\lambda I-\Omega)=\lambda^4+b\lambda^2+c,
\end{equation}
where
\begin{equation}\label{22b}
b=\omega_x^2+\omega_y^2+4\omega_B^2,\quad c=\omega_x^2\omega_y^2.
\end{equation}
\end{mathletters}
Since
\begin{equation}\label{23}
\Delta\equiv
b^2-4c=(\omega_x^2-\omega_y^2)^2+8\omega_B^2(\omega_x^2+\omega_y^2
+2\omega_B^2)\ge 0,
\end{equation}
we have two real roots for $\lambda^2$. If $\omega_x=0$ or
$\omega_y=0$, one of the two roots is zero, and the following
discussions are not valid. We will return to this case latter.
Currently we assume that both $\omega_x$ and $\omega_y$ are nonzero,
then $c>0$ and $\sqrt\Delta<b$, and both roots for $\lambda^2$ are
negative. Therefore the above characteristic polynomial has four pure
imaginary roots:
\begin{equation}\label{24}
\{\lambda_1,\lambda_2,\lambda_3,\lambda_4\}=\{-{\mathrm i}\sigma_1,
{\mathrm i}\sigma_1, -{\mathrm i}\sigma_2, {\mathrm i}\sigma_2\},
\end{equation}
where
\begin{equation}\label{25}
\sigma_1=\left(b+\sqrt\Delta\over 2\right)^{1/2},\quad
\sigma_2=\left(b-\sqrt\Delta\over 2\right)^{1/2},
\end{equation}
and $\sigma_1\ge\sigma_2>0$. The equal sign appears when $\omega_B=0$
and $\omega_x=\omega_y$.

Because $\Omega$ is not symmetric, right eigenvectors and left ones
are different. We define two left eigenvectors (row vectors) $u_i$
corresponding to the eigenvalues $-{\mathrm i}\sigma_i$ ($i=1,2$) by
\begin{equation}\label{26}
u_i\Omega=-{\mathrm i}\sigma_i u_i,\quad i=1,2,
\end{equation}
then the other two are $u_i^*$, corresponding to the eigenvalues
${\mathrm i}\sigma_i$ ($i=1,2$). Similarly, the right eigenvectors
(column vectors) are $v_i$ and $v_i^*$ ($i=1,2$), satisfying the
equations
\begin{equation}\label{27}
\Omega v_i=-{\mathrm i}\sigma_i v_i,\quad i=1,2
\end{equation}
and the complex conjugate ones. From these eigenvalue equations it
can be shown that
\begin{mathletters}\label{28}
\begin{equation}\label{28a}
u_i^* v_j=u_i v_j^*=0, \quad \forall\; i,j=1,2,
\end{equation}
and $u_i v_j=u_i^* v_j^*=0$ for $\sigma_i\ne\sigma_j$ (this should be
checked individually when $\sigma_1=\sigma_2$, see below). By
appropriately choosing the normalization constants we have
\begin{equation}\label{28b}
u_i v_j=u_i^* v_j^*=\delta_{ij}, \quad \forall\; i,j=1,2.
\end{equation}
\end{mathletters}
The specific expression for the above eigenvectors are not necessary
in this section. However, a relation between the left and right
eigenvectors is very useful in the following. Using Eqs. (\ref{20}),
(\ref{26}) and (\ref{27}), it is easy to show that $v_i\propto
\Sigma_y u_i^\dagger$, so we choose
\begin{equation}\label{29}
v_i=-\Sigma_y u_i^\dagger, \quad i=1,2.
\end{equation}
Note that the condition (\ref{28b}) only determine the product of the
normalization constants in $u_i$ and $v_i$. With this relation both
constants are fixed (up to a phase factor). Also note that Eqs.
(\ref{28b}) and (\ref{29}) lead to $-u_i\Sigma_y u_i^\dagger=1$ (no
summation over $i$ on the left-hand side). The left-hand side of this
equation is real (which can be easily shown), but not necessarily
positive. Actually it can be shown by using the eigenvalue equation
that $-u_i\Sigma_y u_i^\dagger=(u_i\Sigma_y){\cal H}
(u_i\Sigma_y)^\dagger/\sigma_i$. Thus the sign of this quantity
depends on the matrix ${\cal H}$, and in general Eq. (\ref{29})
should be replaced by $v_i=\epsilon_i\Sigma_y u_i^\dagger$ where
$\epsilon_i=\pm 1$. For the present case, however, Eq. (\ref{29}) is
sound.

Now we define a $4\times 4$ matrix $Q$ by arranging the column
vectors in the following order:
\begin{equation}\label{30}
Q=(v_1, v_1^*, v_2, v_2^*).
\end{equation}
Using Eq. (\ref{28}) it is easy to show that
\begin{equation}\label{31}
Q^{-1}=(u_1^\tau, u_1^{*\tau}, u_2^\tau, u_2^{*\tau})^\tau.
\end{equation}
With the help of the eigenvalue equations (\ref{26}) and (\ref{27}),
we have
\begin{equation}\label{32}
Q^{-1}\Omega Q={\rm diag}(-{\mathrm i}\sigma_1, {\mathrm i}\sigma_1,
-{\mathrm i}\sigma_2, {\mathrm i}\sigma_2).
\end{equation}
Therefore the matrix $\Omega$ is diagonalized. Using the relation
(\ref{29}), it is not difficult to show that
\begin{equation}\label{33}
Q^\dagger=-\Sigma_z Q^{-1} \Sigma_y,
\end{equation}
where $\Sigma_z={\rm diag}(\sigma_z,\sigma_z)$, and $\sigma_z$ is the
third Pauli matrix. This relation is very useful in the following.

Next we will diagonalize the Hamiltonian. With the above results this
is easy. We define two operators
\begin{mathletters}\label{34}
\begin{equation}\label{34a}
a_i=u_i X/\sqrt\hbar,\quad i=1,2.
\end{equation}
Their hermitian conjugates are
\begin{equation}\label{34b}
a_i^\dagger=u_i^* X/\sqrt\hbar,\quad i=1,2.
\end{equation}
\end{mathletters}
Using Eqs. (\ref{18}), (\ref{28}) and (\ref{29}) it is easy to show
that the nonvanishing commutators among these operators are
\begin{equation}\label{35}
[a_i, a_j^\dagger]=\delta_{ij}, \quad i,j=1,2.
\end{equation}
Therefore they are similar to the raising and lowering operators for
harmonic oscillators. We define a column vector
\begin{equation}\label{36}
A=(a_1, a_1^\dagger, a_2, a_2^\dagger)^\tau,
\end{equation}
then Eq. (\ref{34}) can be written in the matrix form
\begin{equation}\label{34'}\eqnum{$34'$}
A=Q^{-1}X/\sqrt\hbar.
\end{equation}
The inverse is
\begin{equation}\label{37}
X=\sqrt\hbar QA,
\end{equation}
and thus
\begin{equation}\label{38}
X^\tau=X^\dagger=\sqrt\hbar A^\dagger Q^\dagger.
\end{equation}
Substituting these into Eq. (\ref{17}), and using Eqs. (\ref{20}),
(\ref{32}) and (\ref{33}), it is rather easy to show that
\begin{equation}\label{39}
H=\textstyle\frac 12 \hbar A^\dagger\Sigma A,
\end{equation}
where $\Sigma={\rm diag}(\sigma_1, \sigma_1, \sigma_2, \sigma_2)$.
Using the commutation relations (\ref{35}) this becomes
\begin{equation}\label{40}
H=\textstyle\hbar\sigma_1(a_1^\dagger a_1 +\frac 12)+
\hbar\sigma_2(a_2^\dagger a_2 +\frac 12).
\end{equation}
Therefore the Hamiltonian is diagonalized. More specifically, it
becomes the sum of two one-dimensional harmonic oscillator
Hamiltonians. The energy levels are readily available:
\begin{equation}\label{41}
{\cal E}_{n_1 n_2}=\textstyle\hbar\sigma_1(n_1 +\frac 12)+
\hbar\sigma_2(n_2 +\frac 12),\quad n_1, n_2=0,1,2,\ldots.
\end{equation}
When $\omega_B=0$, we have $\sigma_1=\max(\omega_x, \omega_y)$ and
$\sigma_2=\min(\omega_x, \omega_y)$, and these energy levels reduce
to known results, as expected. Substituting these into Eq. (\ref{15})
we obtain the energy levels for the motion on the $xy$ plane:
\begin{equation}\label{42}
{\cal E}^{xy}_{n_1 n_2}=\hbar\sigma_1\left(n_1 +\frac 12\right)+
\hbar\sigma_2\left(n_2 +\frac 12\right)-{q^2 E_x^2 \over
2M\omega_x^2} -{q^2 E_y^2 \over 2M\omega_y^2},\quad n_1,
n_2=0,1,2,\ldots.
\end{equation}
The total energy is the sum of ${\cal E}^{xy}_{n_1 n_2}$ and ${\cal
E}_{n_3}^z$ given by Eq. (\ref{8}). The wave functions on the $xy$
plane will be worked out in the next section.

The method employed in this section is similar to that used in Ref.
\cite{5} which dealt with a charged particle moving in a rotating
magnetic field (also studied in Refs. \cite{6} and \cite{7}).
However, we present here a relation between the right and left
eigenvectors, which leads to the very useful relation (\ref{33}), so
that the calculations here seems more straightforward and easier. The
method can be easily extended to systems with more degrees of
freedom.

\section{The wave functions in the coordinate representation} 

The eigenstates of the Hamiltonian (\ref{40}) corresponding to the
eigenvalues (\ref{41}) are
\begin{equation}\label{43}
|n_1n_2\rangle={1\over \sqrt{n_1! n_2!}}
(a_1^\dagger)^{n_1}(a_2^\dagger)^{n_2}|00\rangle, \quad n_1,
n_2=0,1,2,\ldots,
\end{equation}
where $|00\rangle$ is the ground state satisfying
\begin{equation}\label{44}
a_1|00\rangle=a_2|00\rangle=0.
\end{equation}
The task of this section is to work out these states in the
coordinate representation, that is, the wave functions
\begin{equation}\label{45}
\psi_{n_1n_2}(x,y)=\langle xy|n_1n_2\rangle.
\end{equation}
For this purpose the left eigenvectors are necessary. These are given
by
\begin{mathletters}\label{46}
\begin{equation}\label{46a}
u_i=K_i{\bbox(} -{\mathrm i}
M\sigma_i(\sigma_i^2-\omega_y^2-2\omega_B^2),\;
\sigma_i^2-\omega_y^2,\; M\omega_B(\sigma_i^2+\omega_y^2),\; {\mathrm
i}2\omega_B\sigma_i {\bbox)},\quad i=1,2
\end{equation}
and their complex conjugates, where the normalization constants are
\begin{equation}\label{46b}
K_i=\{{2M\sigma_i}[(\sigma_i^2-\omega_y^2)^2+4\omega_B^2\omega_y^2]\}
^{-1/2}.
\end{equation}
\end{mathletters}
When $\omega_B=0$, $\{\sigma_i|i=1,2\}=\{\omega_x, \omega_y\}$. For
$\omega_x$, the last two components of $u_i$ vanish. For $\omega_y$,
the first two components of $u_i$ vanish. Thus the orthogonality is
ensured. If the further condition $\omega_x=\omega_y$ holds (which
yields $\sigma_1=\sigma_2$), one may take the limit $\omega_y\to
\omega_x$ after everything is worked out.

First we should work out the ground state. In the coordinate
representation, Eq. (\ref{44}) takes the form
\begin{equation}\label{47}
(\xi_{ij}\tilde x_j-{\mathrm
i}\hbar\eta_{ij}\partial_j)\psi_{00}(x,y)=0,\quad i=1,2,
\end{equation}
where $x_1=x$, $x_2=y$ and similarly for $\tilde x_1$ and $\tilde
x_2$, $\partial_j=\partial/\partial x_j=\partial/\partial\tilde x_j$,
and we have defined two $2\times 2$ matrices
\begin{equation}\label{48}
\xi=(\xi_{ij})=\left(
\begin{array}{cc}
  u_{11} & u_{13} \\
  u_{21} & u_{23}
\end{array}
\right),\quad \eta=(\eta_{ij})=\left(
\begin{array}{cc}
  u_{12} & u_{14} \\
  u_{22} & u_{24}
\end{array}
\right),
\end{equation}
where $u_{i\beta}$ ($i=1,2$; $\beta=1,2,3,4$) is the $\beta$th
component of $u_i$. Suppose that
\begin{equation}\label{49}
\psi_{00}(x,y)=N_0\exp[-s(x, y)],\quad s(x, y)={\tilde x_i
\Lambda_{ij} \tilde x_j/2\hbar},
\end{equation}
where $S$ is a $2\times 2$ symmetric matrix whose elements are
complex numbers, and $N_0$ is a normalization constant. Substituting
this into Eq. (\ref{47}) we obtain
\begin{equation}\label{50}
\Lambda={\mathrm i}\eta^{-1}\xi.
\end{equation}
This can be easily worked out. The elements are
\begin{mathletters}\label{51}
\begin{equation}\label{51a}
\Lambda_{11}={M\omega_x(\sigma_1+\sigma_2) \over \omega_x+\omega_y}
\equiv \hbar\lambda_x^2,
\end{equation}
\begin{equation}\label{51b}
\Lambda_{22}={M\omega_y(\sigma_1+\sigma_2) \over \omega_x +\omega_y}
\equiv \hbar\lambda_y^2,
\end{equation}
\begin{equation}\label{51c}
\Lambda_{12}=\Lambda_{21}={{\mathrm i}M\omega_B(\omega_x-\omega_y)
\over \omega_x +\omega_y} \equiv {\mathrm i}\hbar\lambda_{xy},
\end{equation}
\end{mathletters}
where the relation $\sigma_1\sigma_2=\omega_x\omega_y$ has been used.
Therefore the ground-state wave function is
\begin{mathletters}\label{52}
\begin{equation}\label{52a}
\psi_{00}(x,y)=N_0\exp(\textstyle -\frac 12 \lambda_x^2\tilde x^2
-\frac 12 \lambda_y^2\tilde y^2-{\mathrm i}\lambda_{xy}\tilde x\tilde
y),
\end{equation}
where the normalization constant is
\begin{equation}\label{52b}
N_0=\sqrt{\lambda_x\lambda_y\over\pi} =\left[{M(\sigma_1+\sigma_2)
\sqrt{\omega_x\omega_y}\over \pi\hbar(\omega_x+\omega_y)}
\right]^{1/2}.
\end{equation}
\end{mathletters}
When $\omega_B=0$, we have $\lambda_x=\sqrt{M\omega_x/\hbar}$,
$\lambda_y=\sqrt{M\omega_y/\hbar}$, and $\lambda_{xy}=0$. On the
other hand, if $\omega_x=\omega_y$, we have $\omega_1=\omega_2$,
$\sigma_1=\omega_1+|\omega_B|$, $\sigma_2=\omega_1-|\omega_B|$, and
$\lambda_x=\lambda_y=\sqrt{M\omega_1/\hbar}$, $\lambda_{xy}=0$. These
are all expected results.

It is remarkable that the uncertainty relation for the above ground
state is
\begin{equation}\label{53}
\Delta x\Delta p_x=\Delta y\Delta p_y={\hbar\over 2}\left(1+
{\lambda_{xy}^2\over\lambda_x^2\lambda_y^2}\right)^{1/2}\ge
{\hbar\over 2}.
\end{equation}
The equal sign holds only when $B=0$ or $\omega_x=\omega_y$.
Therefore the ground state is in general not a minimum wave packet,
though it is of the Gaussian type. It seems that this is not noticed
in the previous literature.

For any function $F(x,y)$, it is easy to show that
\begin{equation}\label{54}
(\xi^*_{ij}\tilde x_j-{\mathrm i}\hbar\eta^*_{ij}\partial_j)F(x,y)
=-{\mathrm i}\hbar\exp[s^*(x,y)]\eta^*_{ij}\partial_j
\{\exp[-s^*(x,y)]F(x,y)\},\quad i=1,2.
\end{equation}
According to Eq. (\ref{43}), and using the above relation we obtain
\begin{eqnarray}\label{55}
\psi_{n_1n_2}(x,y)=&&{(-{\mathrm i}\sqrt\hbar)^{n_1+n_2}
\over\sqrt{n_1!n_2!}} N_0K_1^{n_1}K_2^{n_2} \exp(\textstyle \frac 12
\lambda_x^2\tilde x^2+ \frac 12 \lambda_y^2\tilde y^2-{\mathrm i}
\lambda_{xy}\tilde x\tilde y)
\nonumber\\
&&[(\sigma_1^2-\omega_y^2)\partial_x -{\mathrm i}
2\omega_B\sigma_1\partial_y] ^{n_1}
[(\sigma_2^2-\omega_y^2)\partial_x -{\mathrm i}
2\omega_B\sigma_2\partial_y] ^{n_2} \exp( -\lambda_x^2\tilde
x^2-\lambda_y^2\tilde y^2).
\end{eqnarray}

\section{Some special cases} 

In this section we discuss some special cases where some of the
parameters are zero.

1. If $\omega_z=0$ and $E_z=0$, the motion in the $z$ direction is
free, and ${\cal E}_z\ge 0$ is continuous.

2. If $\omega_z=0$ but $E_z\ne 0$, the motion in the $z$ direction is
that in a homogeneous electric field, which is also well known, and
${\cal E}_z$ is also continuous, and can take on any real value
\cite{8}.

3. In the preceding sections we have assumed that both $\omega_x$ and
$\omega_y$ are nonzero. In this case, if any one of $E_x$ or $E_y$ is
zero, it can be substituted into the above results directly and no
modification to the formalism is needed. However, if one of
$\omega_x$ and $\omega_y$ is zero, the situation is different. We
would deal with $\omega_y=0$ in the following. The other case could
be discussed in a similar way.

4. If $\omega_y=0$ but $E_y\ne 0$, the Hamiltonian could not be
reduced to a quadratic form, and to our knowledge the problem could
not be solved analytically.

5. If $\omega_y=0$ and $E_y=0$, we choose the gauge $A_x=0$,
$A_y=Bx$, and let
\begin{equation}\label{56}
\psi(x,y)={1\over\sqrt{2\pi}}\exp({\mathrm i}ky)\phi(x),
\end{equation}
then $\phi(x)$ satisfies the equation
\begin{equation}\label{57}
\left[{1 \over 2M}p_x^2 +{1\over 2}M\tilde\omega_1^2 (x-x_k)^2\right]
\phi(x) ={\cal E}^x\phi(x),
\end{equation}
where
\begin{equation}\label{58}
\tilde\omega_1=\sqrt{\omega_x^2+4\omega_B^2},\quad
x_k={qE_x+2k\hbar\omega_B \over M\tilde\omega_1^2},
\end{equation}
and
\begin{equation}\label{59}
{\cal E}^x={\cal E}^{xy}+{(qE_x+2k\hbar\omega_B)^2 \over
2M\tilde\omega_1^2}-{\hbar^2 k^2 \over 2M}.
\end{equation}
The energy levels and wave functions for Eq. (\ref{57}) are readily
available, so the final results are
\begin{equation}\label{60}
{\cal E}^{xy}_{n_1 k}=\hbar\tilde\omega_1\left(n_1 +\frac 12\right)-
{(qE_x+2k\hbar\omega_B)^2 \over 2M\tilde\omega_1^2}+{\hbar^2 k^2
\over 2M},\quad n_1=0,1,2,\ldots,\; k\in(-\infty,+\infty),
\end{equation}
and the corresponding wave functions are
\begin{equation}\label{61}
\psi_{n_1 k}(x,y)={1\over\sqrt{2\pi}}\exp({\mathrm i}ky)
\psi_{n_1}^{\tilde\omega_1}(x-x_k).
\end{equation}
If further $\omega_x$ or $E_x$ or both are zero, the results can be
obtained from the above ones directly.

6. Finally, if both $\omega_x$ and $\omega_y$ are zero, one can
appropriately choose the coordinate axes of the $xy$ plane such that
$E_y=0$. Thus it is a special case of the above case 5.

\section{A more general magnetic field}  

In this section we set $\omega_z=0$ and $E_z=0$, and consider a
magnetic field with an $x$ or $y$ component (but not both) in
addition to the $z$ component. To our knowledge this was not
considered previously. Without loss of generality we take
\begin{equation}\label{62}
{\mathbf B}=B'{\mathbf e}_x+B{\mathbf e}_z.
\end{equation}
The gauge convenient for the present case might be
\begin{equation}\label{63}
A_x=0,\quad A_y=B\tilde x=B(x-x_0),\quad A_z=B'y,
\end{equation}
where $x_0$ is the same as defined before. In this gauge the total
Hamiltonian takes the form
\begin{equation}\label{64}
H^{\rm T}={1 \over 2M}(p_x^2 +p_y^2) +{1 \over
2M}(p_z-2M\omega_B'y)^2 +{1\over 2}M\tilde\omega_1^2\tilde x^2
+{1\over 2}M\omega_y^2 y^2 -qE_y y- 2\omega_B\tilde xp_y-{q^2 E_x^2
\over 2M\omega_x^2},
\end{equation}
where $\omega_B$ and $\tilde\omega_1$ are the same as defined before,
and $\omega_B'=qB'/2Mc$. $p_z$ is obviously a conserved quantity, so
we make the factorization
\begin{equation}\label{65}
\Psi({\mathbf x}) ={1\over\sqrt{2\pi}}\exp({\mathrm i}kz)\psi(x,y).
\end{equation}
The Schr\"odinger equation (\ref{1}) is then reduced to the form
(\ref{14a}) where now
\begin{equation}\label{66}
H={1 \over 2M}(p_x^2 +p_y^2) +{1\over 2}M\tilde\omega_1^2 \tilde x^2
+{1\over 2}M\tilde\omega_2^2 \tilde y^2-2\omega_B\tilde xp_y,
\end{equation}
and
\begin{equation}\label{67}
{\cal E}={\cal E}^{\rm T}+{q^2E_x^2 \over
2M\omega_x^2}+{(qE_y+2k\hbar\omega_B')^2 \over
2M\tilde\omega_2^2}-{\hbar^2 k^2 \over 2M}.
\end{equation}
In these equations $\tilde y=y-y_k$ which is different from the
previous one, and
\begin{equation}\label{68}
\tilde\omega_2=\sqrt{\omega_y^2+4\omega_B'^2},\quad
y_k={qE_y+2k\hbar\omega_B' \over M\tilde\omega_2^2}.
\end{equation}
Now the reduced Hamiltonian $H$ is quadratic, and different from Eq.
(\ref{14b}) only in the last term. Thus the problem can be solved in
much the same way as before. We only give the results here. The
energy levels are
\begin{eqnarray}\label{69}
{\cal E}^{\rm T}_{n_1 n_2 k}=&&\hbar\sigma_1\left(n_1 +\frac
12\right)+ \hbar\sigma_2\left(n_2 +\frac 12\right)-{q^2 E_x^2 \over
2M\omega_x^2} -{(qE_y+2k\hbar\omega_B')^2 \over
2M\tilde\omega_2^2}+{\hbar^2 k^2 \over 2M},\nonumber\\ &&  n_1,
n_2=0,1,2,\ldots, \; k\in(-\infty,+\infty),
\end{eqnarray}
where $\sigma_1$ and $\sigma_2$ are given by Eqs. (\ref{25}),
(\ref{22b}) and (\ref{23}), but with $\omega_y$ replaced by
$\tilde\omega_2$. The wave functions are given by
\begin{equation}\label{70}
\Psi_{n_1n_2k}({\mathbf x}) ={1\over\sqrt{2\pi}}\exp({\mathrm i}kz)
\psi_{n_1n_2k}(x,y),
\end{equation}
where $\psi_{n_1n_2k}(x,y)$ has the same form as Eq. (\ref{55}), with
$\omega_y$ replaced by $\tilde\omega_2$ everywhere, including in the
expressions for $\lambda_x$, $\lambda_y$, $K_1$ and $K_2$, but here
\begin{equation}\label{71}
\lambda_{xy}=-{2M\omega_B\tilde\omega_2 \over \hbar(\omega_x
+\tilde\omega_2)}.
\end{equation}
Also note that $\tilde y$ depends on $k$, which is the reason why
$\psi_{n_1n_2k}(x,y)$ has the subscript $k$. The orthonormal relation
is
\begin{equation}\label{72}
\int \Psi^*_{n_1'n_2'k'}({\mathbf x})\Psi_{n_1n_2k}({\mathbf x})\;
d{\mathbf x} =\delta(k-k')\delta_{n_1 n_1'}\delta_{n_2 n_2'}.
\end{equation}
By the way, the wave functions (\ref{61}) satisfy a similar
orthonormal relation.

As before, the ground state obtained here, though being of the
Gaussian type, is not a minimum wave packet, except when $B=0$ (then
${\mathbf B}=B'{\mathbf e}_x$).

\section{Coherent and squeezed states}  

Coherent and squeezed states are useful objects in quantum mechanics
and quantum optics. These are widely discussed in the literature. In
the presence of magnetic fields, some discussions can be found in
Ref. \cite{fan}. There are also examples on other applications of
such states \cite{jing}. However, such states for anisotropic
harmonic oscillators in the presence of magnetic fields are not
discussed previously, to the best of our knowledge. In this section
we will discuss the definition, the properties and the time evolution
of such states for the case treated in Sec. III and IV. The
discussions can be easily extended to the case of Sec. VI.

Since the motion in the $z$ direction is separable and is essentially
that of a usual harmonic oscillator, we only discuss the coherent and
squeezed states on the $\tilde x\tilde y$ plane (which is a simple
translation of the $xy$ plane), and their time evolution as governed
by the Hamiltonian $H$ given by Eq. (\ref{14a}) or (\ref{40}) (the
additional constants in $H^{xy}$ has only trivial consequence for the
time evolution). The definitions used below are natural
generalizations of those for a single harmonic oscillator
\cite{nieto,ni,ka}.

We define a unitary displacement operator
$D(\alpha_1,\alpha_2)=D_1(\alpha_1) D_2(\alpha_2)$ where
\begin{equation}\label{73}
D_i(\alpha_i)=\exp(\alpha_i a_i^\dagger-\alpha_i^* a_i),\quad i=1,2
\end{equation}
and the $\alpha_i$'s are complex numbers. For an arbitrary state
$|\varphi\rangle$ one can define a corresponding displaced state
\begin{equation}\label{74}
|\varphi,\alpha_1\alpha_2\rangle_D=D(\alpha_1,\alpha_2)|\varphi\rangle=
D_1(\alpha_1) D_2(\alpha_2)|\varphi\rangle.
\end{equation}
It is easy to show that
\begin{equation}\label{75}
D^\dagger(\alpha_1,\alpha_2)a_i D(\alpha_1,\alpha_2)=a_i +\alpha_i,
\quad i=1,2,
\end{equation}
so if $|\varphi\rangle=|00\rangle$ is the ground state we have
\begin{equation}\label{76}
a_i|00,\alpha_1\alpha_2\rangle_D=\alpha_i|00,\alpha_1\alpha_2\rangle_D,
\quad i=1,2.
\end{equation}
Thus the coherent states may be defined as the displaced ground state
$|00,\alpha_1\alpha_2\rangle_D$. In terms of the original variables
the displacement operator takes the form
\begin{equation}\label{77}
D(\alpha_1,\alpha_2)=\exp\left(-{{\mathrm i}\over
2\hbar}x^D_{i}p^D_{i}\right) \exp\left({{\mathrm i}\over
\hbar}p^D_{i}\tilde x_i\right) \exp\left(-{{\mathrm i}\over
\hbar}x^D_{i}p_{i}\right),
\end{equation}
where
\begin{equation}\label{78}
x^D_i={\mathrm i}\sqrt\hbar(\alpha_j\eta^*_{ji}-{\mathrm c.c.}),
\quad p^D_i=-{\mathrm i}\sqrt\hbar(\alpha_j\xi^*_{ji}- {\mathrm
c.c.}).
\end{equation}
If the wave function for the state $|\varphi\rangle$ is
$\varphi(\tilde x,\tilde y)$, then the one for the displaced state
$|\varphi,\alpha_1\alpha_2\rangle_D$ is
\begin{equation}\label{79}
\varphi_{\alpha_1\alpha_2}^D(\tilde x,\tilde y)=\exp\left(-{{\mathrm
i}\over 2\hbar}x^D_{i}p^D_{i}\right) \exp\left({{\mathrm i}\over
\hbar}p^D_{i}\tilde x_i\right) \varphi(\tilde x-x^D,\tilde y-y^D).
\end{equation}
Thus we see the reason why the operator $D(\alpha_1,\alpha_2)$ is
called a displacement operator. From this and Eq. (\ref{52}) it is
easy to obtain the wave function for the coherent states. Obviously
the displaced state and the original one have the same shape in the
configuration space, thus the uncertainty $\Delta X_\alpha$ in
$|\varphi,\alpha_1\alpha_2\rangle_D$ is the same as that in
$|\varphi\rangle$. This can also be easily shown by using Eqs.
(\ref{37}) and (\ref{75}).

Now consider the time evolution of the displaced states. It is easy
to show that
\begin{equation}\label{80}
{\mathrm e}^{-{\mathrm i}Ht/\hbar}a_i {\mathrm e}^{{\mathrm i}Ht
/\hbar}=\exp({\mathrm i}\sigma_i t)a_i, \quad i=1,2.
\end{equation}
If the state at the initial time $t=0$ is $|\psi(0)\rangle=
|\varphi,\alpha_1\alpha_2 \rangle_D$, then the state at the time $t$
is
\begin{equation}\label{81}
|\psi(t)\rangle=|\varphi(t),\alpha_{1t}\alpha_{2t}\rangle_D
=D_1(\alpha_{1t})D_2(\alpha_{2t}) |\varphi(t)\rangle,
\end{equation}
where $\alpha_{it}=\exp(-{\mathrm i}\sigma_i t)\alpha_i$ and
$|\varphi(t)\rangle= {\mathrm e}^{-{\mathrm i}Ht/\hbar}
|\varphi\rangle.$ Therefore if $|\varphi(t)\rangle$ is known,
$|\psi(t)\rangle$ can be obtained by a time-dependent displacement. A
simple special case is $|\varphi\rangle=|n_1 n_2\rangle$, that is
\begin{equation}\label{82}
|\psi(0)\rangle=|n_1 n_2,\alpha_1\alpha_2 \rangle_D,
\end{equation}
a displaced number state. In this case
$|\varphi(t)\rangle=\exp[-{\mathrm i}(n_1+\textstyle\frac 12)\sigma_1
t -{\mathrm i}(n_2+\textstyle\frac 12)\sigma_2 t]|n_1 n_2\rangle$,
and
\begin{equation}\label{83}
|\psi(t)\rangle=\exp[-{\mathrm i}(n_1+\textstyle\frac 12)\sigma_1 t
-{\mathrm i}(n_2+\textstyle\frac 12)\sigma_2 t] |n_1 n_2,
\alpha_{1t}\alpha_{2t} \rangle_D.
\end{equation}
This means that $|\psi(t)\rangle$ is also a displaced number state,
except that the displacement parameters are time dependent, and a
time-dependent phase factor is gained. The position and velocity of
the wave packet $\psi(t,\tilde x,\tilde y)=\langle \tilde x\tilde
y|\psi(t)\rangle$ is characterized by the expectation values
$x_\alpha^{\mathrm c}(t)= \langle\psi(t)|X_\alpha|\psi(t)\rangle$.
Using Eq. (\ref{37}) it is easy to show that
\begin{equation}\label{84}
x_\alpha^{\mathrm c}(t)=\sqrt\hbar[\exp(-{\mathrm i}\sigma_j t)
\alpha_j v_{j\alpha}+{\mathrm c.c.}].
\end{equation}
By straightforward calculations one can show that they satisfy the
following equation
\begin{equation}\label{85}
\dot x_\alpha^{\mathrm c}(t)=\Omega_{\alpha\beta} x_\beta^{\mathrm
c}(t),
\end{equation}
where the eigenvalue equation (\ref{27}) has been used, and at $t=0$
they give the same results as those obtained by using Eqs. (\ref{78})
and (\ref{79}). On the other hand, the equations of motion for
$X_\alpha$ governed by the Hamiltonian $H$ according to classical
mechanics are
\begin{equation}\label{86}
\dot X_\alpha=\{X_\alpha,H\}_{\mathrm PB}=\Omega_{\alpha\beta}
X_\beta,
\end{equation}
where the Poisson bracket $\{X_\alpha,H\}_{\mathrm PB}$ turns out to
be of the same form as the commutator $[X_\alpha,H]/{\mathrm i}\hbar$
because $H$ is quadratic in $X_\alpha$. The above two equations mean
that the center of the wave packet moves like a classical particle.
In fact, it can be shown that this holds for any wave packet in any
quadratic system. For the displaced number states, some more specific
properties should be emphasized. First, the shape of the wave packet
keep unchanged with time. Second, the center of the wave packet is
oscillating, but the motion is in general not periodic except when
$\sigma_2/\sigma_1$ is a rational number.

Next we discuss the squeezed states. We define a unitary squeeze
operator $S(\zeta_1,\zeta_2)=S_1(\zeta_1) S_2(\zeta_2)$ where
\begin{equation}\label{87}
S_i(\zeta_i)=\exp(\textstyle\frac12\zeta_i a_i^\dagger a_i^\dagger
-\frac12\zeta_i^* a_i a_i),\quad i=1,2,
\end{equation}
and the $\zeta_i$'s are complex numbers. For an arbitrary state
$|\varphi\rangle$ the corresponding squeezed state may be defined as
\begin{equation}\label{88}
|\varphi,\zeta_1\zeta_2\rangle_S=S(\zeta_1,\zeta_2)|\varphi\rangle=
S_1(\zeta_1) S_2(\zeta_2)|\varphi\rangle.
\end{equation}
The following equation is useful for subsequent calculations.
\begin{equation}\label{89}
S^\dagger(\zeta_1,\zeta_2)a_i S(\zeta_1,\zeta_2)=a_i\cosh\rho_i
+a_i^\dagger{\mathrm e}^{{\mathrm i}\phi_i}\sinh\rho_i, \quad i=1,2,
\end{equation}
where $\rho_i=|\zeta_i|$ and $\phi_i=\arg\zeta_i$.

Unlike the displaced state, it is difficult to obtain the explicit
wave function for the squeezed state in terms of the wave function
for the original one. However, in the squeezed number state $|n_1n_2,
\zeta_1 \zeta_2 \rangle_S$, it can be shown by using Eq. (\ref{89})
that
\begin{equation}\label{90}
\langle X_\alpha\rangle=0,
\end{equation}
\begin{equation}\label{91}
\Delta X_\alpha=\{\hbar(2n_j+1)[|v_{j\alpha}|^2\cosh2\rho_j+{\mathrm
Re}(v_{j\alpha}^2{\mathrm e}^{{\mathrm i}\phi_j})\sinh2\rho_j]\}
^{1/2}.
\end{equation}
Compared with the results for $\zeta_1=\zeta_2=0$, we see that the
center of the squeezed state is the same as that of the original one,
but the uncertainties are different. Thus the states are indeed
``squeezed''.

Now consider the time evolution of the squeezed states. If the state
at the initial time $t=0$ is $|\psi(0)\rangle=
|\varphi,\zeta_1\zeta_2 \rangle_S$, then the state at the time $t$ is
\begin{equation}\label{92}
|\psi(t)\rangle=|\varphi(t),\zeta_{1t}\zeta_{2t}\rangle_S
=S_1(\zeta_{1t})S_2(\zeta_{2t}) |\varphi(t)\rangle,
\end{equation}
where $\zeta_{it}=\exp(-{\mathrm i}2\sigma_i t)\zeta_i$ and
$|\varphi(t)\rangle= {\mathrm e}^{-{\mathrm i}Ht/\hbar}
|\varphi\rangle.$ Therefore if $|\varphi(t)\rangle$ is known,
$|\psi(t)\rangle$ can be obtained by a time-dependent squeeze. A
simple special case is $|\varphi\rangle=|n_1 n_2\rangle$, that is
\begin{equation}\label{93}
|\psi(0)\rangle=|n_1 n_2,\zeta_1\zeta_2 \rangle_S,
\end{equation}
a squeezed number state. In this case
$|\varphi(t)\rangle=\exp[-{\mathrm i}(n_1+\textstyle\frac 12)\sigma_1
t -{\mathrm i}(n_2+\textstyle\frac 12)\sigma_2 t]|n_1 n_2\rangle$,
and
\begin{equation}\label{94}
|\psi(t)\rangle=\exp[-{\mathrm i}(n_1+\textstyle\frac 12)\sigma_1 t
-{\mathrm i}(n_2+\textstyle\frac 12)\sigma_2 t] |n_1 n_2,
\zeta_{1t}\zeta_{2t} \rangle_S.
\end{equation}
This means that $|\psi(t)\rangle$ is also a squeezed number state,
except that the squeeze parameters are time dependent, and a
time-dependent phase factor is gained. Though it is difficult to
obtain the wave function $\psi(t,\tilde x,\tilde y)=\langle \tilde
x\tilde y|\psi(t)\rangle$ explicitly, in these states it can be shown
that
\begin{equation}\label{95}
\langle X_\alpha\rangle_t=0,
\end{equation}
and
\begin{equation}\label{96}
\Delta_t
X_\alpha=\{\hbar(2n_j+1)[|v_{j\alpha}|^2\cosh2\rho_j+{\mathrm
Re}(v_{j\alpha}^2{\mathrm e}^{{\mathrm i}\phi_j-{\mathrm i}2\sigma_j
t})\sinh2\rho_j]\} ^{1/2}.
\end{equation}
These results mean that the center of the wave packet is at rest, but
the uncertainties are oscillating. Thus the shape of the wave packet
changes with time apparently. As before, the motion is in general not
periodic except when $\sigma_2/\sigma_1$ is a rational number.

\section{Summary}  
In this paper we have studied a charged anisotropic harmonic
oscillator moving in a homogeneous static electromagnetic field.
Several configurations of the electromagnetic field are considered.
One of these configurations has been studied in the literature.
However, the formalism used here seems more convenient. We have
studied the coherent and squeezed states of these systems in some
detail. In terms of these concepts the time evolution of some wave
packets can be discussed very conveniently.

\section*{Acknowledgments}

This work was supported by the National Natural Science Foundation of
the People's Republic of China, and by the Foundation of the Advanced
Research Center of Sun Yat-Sen University.

\section*{Note added after publication}

After this paper had been published we became aware of some more
previous works dealing with the same problem \cite{Schuh,Davies}. In
Ref. \cite{Schuh} the energy spectrum was worked out by an algebraic
approach. In Ref. \cite{Davies} the propagator was calculated by a
different method from that of Refs. \cite{3,4}. We thank Prof. J. M.
Cervero for pointing out these papers to us.


\end{document}